# FASHION AND ART CYCLES ARE DRIVEN BY COUNTER-DOMINANCE SIGNALS OF ELITE COMPETITION: QUANTITATIVE EVIDENCE FROM MUSIC STYLES

Peter Klimek[1,2], Robert Kreuzbauer[3], Stefan Thurner[1,2,4,5]

[1]Section for Science of Complex Systems, CeMSIIS, Medical University of Vienna, Spitalgasse 23, A-1090 Vienna, Austria

[2]Complexity Science Hub Vienna, Josefstaedter Strasse 39, A-1080 Vienna, Austria

[3]University of Surrey, Surrey Business School, Department of Marketing and Retail Management, Guildford, Surrey, GU2 7XH, United Kingdom

[4]Santa Fe Institute, 1399 Hyde Park Road, Santa Fe, NM 85701, USA

[5]IIASA, Schlossplatz 1, A-2361 Laxenburg, Austria

**Keywords**: cultural evolution; evolutionary dynamics; network analysis; fashion cycle theory

## Abstract

Human symbol systems such as art and fashion styles emerge from complex social processes that govern the continuous re-organization of modern societies. They provide a signaling scheme that allows members of an elite to distinguish themselves from the rest of society. Efforts to understand the dynamics of art and fashion cycles have been based on 'bottom-up' and 'top down' theories. According to 'top down' theories, elite members signal their superior status by introducing new symbols (e.g., fashion styles), which are adopted by low-status groups. In response to this adoption, elite members would need to introduce new symbols to signal their status. According to many 'bottom-up' theories, style cycles evolve from lower classes and follow an essentially random pattern. We propose an alternative explanation based on counter-dominance signaling. There, elite members want others to imitate their symbols; changes only occur when outsider groups successfully challenge the elite by introducing signals that contrast those endorsed by the elite. We investigate these mechanisms using a dynamic network approach on data containing almost 8 million musical albums released between 1956 and 2015. The network systematically quantifies artistic similarities of competing musical styles and their changes over time. We formulate empirical tests for whether new symbols are introduced by current elite members (top-down), randomness (bottom-up) or by peripheral groups through counter-dominance signals. We find clear evidence that counter-dominance-signaling drives changes in musical styles. This provides a quantitative, completely data-driven answer to a century-old debate about the nature of the underlying social dynamics of fashion cycles.

## Introduction

Of all species, only modern homo sapiens has evolved the ability to use complex symbolic behavior to organize and coordinate large anonymous societies [1,2,3]. Symbols facilitate the identification of group membership, social status, and, consequently, the competition between elites—here defined as social groups, which have disproportionate access of control over economic, social, cultural, political, or knowledge resources [4,5]. In recent times, 'symbolic social coordination and competition' are perhaps best captured in fashion and art style cycles (e.g. in music, literature, architecture) [6,7,8,9,12]. There, the social influence of an elite is represented by the number of people who adopt the elite's core stylistic elements [10,11]. *Costly signaling theory* (ST) has been proposed as an explanation for the social and cultural evolutionary dynamics that determine the changes of fashion and art styles [3,13]. According to ST, elite members introduce a style which is too costly to adopt for non-members (e.g., a chandelier) and hence provides an honest signal of the elite's superior status. Yet, non-elite members would sooner or later find ways to mimic the elite style (e.g., acquiring inexpensive chandeliers). Elite members would then abandon the style and replace it with a new one (e.g. a modern-style functionalistic pendant light), which again is too costly to adopt for non-members [6]; they are constantly "on the run" from too many adopters and innovate in reaction to being imitated. Often the new style of the elite would be in stark contrast to the previous style, as shifting to a style highly distinctive from the previous one adds an additional obstacle for the adoption by lower classes [14]. This would explain why consecutive styles often occur as extreme opposites (e.g., ornamentation versus functionalism). However, this 'top-down' mechanism (stylistic elements spread from elite to non-members) has been questioned by others who propose a 'bottom-up' dynamic (elite members adopt stylistic elements from non-members to retain their status) to account for the fact that styles often evolve from lower classes or sub-cultures (e.g., Punk-rock or ripped jeans) [8,15,16,17]. Following this logic, broad stylistic changes are driven by external factors such as symbolic elements chosen and promoted by cultural industries (e.g., trend-scouts from fashion houses or music labels who may just randomly pick up symbols from a sub-culture) [18,19,20]. According to this view, fashion cycles should follow an essentially random pattern. This has been formalized in the so-called *Random-Pattern-Theories* (RPT). For instance, it has been shown that models based on the random copying of cultural traits can explain cyclical patterns of fashion change in a quite robust way, for instance in the popularity of first names [21, 22, 23]. The central assumption in these models is that some styles will become highly popular simply due to imitation but not because they are in some way superior to other styles. However, there exist problems with both the ST and the RPT approach. For example, various 'elite-styles' enjoy broad adoption by the middle and lower classes (e.g., ties, black suits, diamond rings) but elite members did not abandon them (as ST would predict). RPT, on the other hand, is unable to account for the fact that two subsequent styles often endorse completely opposite sets of symbols, such as the change from modernism to postmodernism. Instead, we argue that such prominent and enduring cyclical symbolic patterns of mass-culture can only be understood by considering the structure of the networks that underlie the processes of social coordination, such as elite competition in the form of counter-dominance signaling.

To resolve the puzzle of conflicting predictions suggested by ST and RPT, we propose a third mechanism, that of *counter-dominance-signaling* (CDS). This mechanism is rooted in cultural evolutionary theories of counter-dominance as well as theories of collective action and social movement [24,25,26,27]. Rather than assuming that members of an elite are constantly on the run from too many adopters, in the CDS framework, elite members actually *like* to see others adopt their symbols and styles. This signals the cultural influence of elite members [10,24]. Stylistic changes occur when a new upcoming elite successfully challenges the dominance of the established elite by introducing a novel style, which is subsequently adopted by a sufficiently large number of followers. In other words, followers stop adopting the style from the established elite A and start to adopt the style from a new emerging elite B. The new style of elite B should starkly deviate from the previous style, since it signals and emphasizes its disagreement with the previous hegemony. Examples for such CDS include ostentatiously decorative and non-functional ornaments of post-modern architecture that have the potential to provocatively signal the protest against the dominant elite group such as the 'form-follows-function' doctrine of modernism [24]. CDS serves as a focal point to facilitate the coordination of people that share a reservation against the current dominant elite group [10,24,26].

Our proposed mechanism of CDS allows us to solve a century old debate on the driving mechanisms behind stylistic changes in areas of mass-culture such as art and fashion [6]. On the one hand—consistent with RPT but inconsistent with ST—CDS is able to account for the fact that new styles are not necessarily introduced by elites, but often do evolve through a 'bottom-up' dynamic where elite members adopt specific stylistic elements from non-members. On the other hand—consistent with ST but inconsistent with RPT—CDS predicts patterns of starkly contrasting symbols in consecutive styles. The contrasts arise because outsiders use highly distinctive symbols to counter the dominance of existing elite groups and not because the elite introduced distinctive elements to prevent the masses from adopting them, as predicted by ST.

A fundamental challenge is to empirically test (i) if non-trivial mechanisms are indeed necessary to understand cultural change and, if so, (ii) which mechanism—ST, RPT, or CDS—is actually realized in society. Therefore, we first formulate a null model for the evolution of musical styles that assumes that music producers choose the styles which they adopt completely at random; this provides a neutral model for the evolution of styles [28]. We show that such a model is not able to account for the existence of fashion cycles observed in the data of musical styles, which suggests that the evolution of styles is driven by mechanisms that depend on stylistic differences. The neutral model can therefore be modified to include the mechanisms of ST, RPT, or CDS, which then indeed leads to the emergence of fashion cycles. Empirical tests are then needed to determine which model mechanism best describes the actual evolution of musical styles. To this end we developed a method to quantify musical styles by determining each style's typical instrumentation. From a dataset containing almost eight million albums that have been released since 1950, we extracted information about a user-created taxonomy of fifteen musical genres, 422 musical styles, and 570 different instruments. The instruments that are typically associated with a given genre (or style) were shown to be a suitable approximation to formally describe the characteristics of a style [29]. Therefore, the similarity between styles can be quantified through the similarity of their instrumentation. For instance, in Figure 1A we show an example of four different musical styles (blue circles) that are linked to five instruments (green squares). Here a link indicates that the instrument is (typically) featured in a release belonging to that style. The higher the overlap in instruments between two styles, the higher is their similarity and the thicker is the line that connects the styles in Figure 1A. Using this network representation, we are able to rephrase the question which mechanism—ST, RPT, or CST—is realized to a question of detecting specific patterns of the network's evolution, see Figure 1B. There, we schematically show three different scenarios for the time evolution of a network (blue circles) that represent the three theories ST, RPT, or CDS, respectively. The size of the circles indicates the popularity of a given style, i.e., how many artists adopted that particular style. Initially, there exists an elite, whose members have broadly adopted style *i*. Other styles with lower popularity (non-elite groups) will imitate the style of this elite group, resulting in an increased similarity of other styles with *i* (increase of thickness of the lines connecting them to *i* in Figure 1B). According to ST, elite members will react to this imitation by adopting a new and different style *j*. Now it will be style *j* that the low-popularity groups seek to follow and imitate. This results in increased popularity of *j* and increased similarity between *j* and the prior elite style *i*. In the RPT scenario, in contrast, it is a new elite that forms at a random position in the network that dominates the next fashion cycle. Finally, the mechanism of CDS suggests that the current elite *i* will be provocatively challenged by an outsider group, *k*, using signals (in our case sets of instruments) that are in stark opposition to those endorsed by *i*. Through the adoption of these signals that oppose *i*, a new counter-elite emerges with style *k*. Each of the three mechanisms leads to the same generic cyclic patterns of fashion styles, see Figure 1C. That is, initially style *i* increases in popularity as it is imitated by other styles. With the shift of the elite to a different style (ST) or the emergence of a random new (RPT) or counter elite (CDS), new styles gain popularity at the expense of style *i* and will therefore more frequently be imitated. This imitation triggers the next fashion cycle. By cycles we do not refer to patterns where a single style oscillates in popularity over time, but rather a pattern where certain styles rise sharply in popularity for some time, eventually level off in their growth, and then fade back into oblivion as new styles emerge that trigger the next cycle. With the help of the proposed network formalism we can specify concrete hypotheses to test whether the observed changes in similarities and popularities are best described by a network evolution mechanism that is compatible with ST, RPT, or CDS.

## Data and Methods

**Data.** Discogs is a crowdsourced database of information about audio releases [30]. As of January 2016, it contained entries on 7,911,789 albums released between 1952 and 2015. Each album is assigned one or several musical styles. In total there are 422 musical styles grouped into 15 genres. For each album, Discogs provides a structured list of credits. These credits include a list of artists and the instruments the artists used in the recording. There are 570 different instruments in the database, ranging from lead vocals, finger snaps and countertenor over drums, electric guitars, keyboards, and violins, to more exotic instruments like hunting horns, Northumbrian pipes, or MIDI controllers. This data structure allows us to characterize each style by a unique combination of instruments used in recordings associated with that style. Styles and instruments are introduced independently from each other in the data. Discogs employs a moderation system to categorize styles into taxonomic hierarchies in a systematic way [30]. Styles cannot be freely introduced by users, instead this requires a certain number of releases, justification for why they are different from currently existing styles, and at least three trustworthy external citations of the style's use. It was statistically confirmed that genres indeed correspond to clusters of mutually similar styles [29]. This means that the way in which instruments are assigned to styles is far from random and highly correlated with the independently obtained folksonomic classification of styles. The emergence of a new style in the data therefore means that a stylistic change has occurred that convinced a large-enough group of experts in popular music to introduce a new category to label this kind of music.

**Style popularity.** Popularity of a style *s* is related to the number of albums released within a given time interval of time *t*, in style *s*, $N_s(t)$. The total number of albums released per year increased from 3,051 in 1952 to more than 200,000 after 2012. To account for these large variations, we define the popularity of a style as the relative frequency of releases within a specific time interval, $n_s(t) = N_s(t)/\sum_s N_s(t)$. This notion of popularity quantifies how likely a music producer is to adopt a given style, which is not necessarily the same as the popularity of the style among music consumers. Popularity change of a style, $\delta n_s(t)$, is measured as the percent change in popularity between year *t* and $t + \delta t$.

**Highs, Lows, and Newcomer.** At each time interval we consider the set of *m* styles with the lowest and highest popularities, to which we refer as *Lows* and *Highs*, respectively. *Newcomer* are defined as those *Lows* that show the highest increases in popularity. That is, we rank styles with at least one release according to their popularity and obtain their popularity rank, $Rank(n_s(t))$. Low (high) rank values indicate high (low) popularity (rank 1 means highest popularity). For each time interval *t* we identify high-popularity styles as follows. *Highs* are all styles with a popularity rank, $Rank(n_s(t))$, below *m*,

$$Highs(t, m) = \{s | Rank(n_s(t)) \leq m\}. \tag{1}$$

*Lows* include the remaining styles that have a popularity rank higher than *m*. In general, the number of such styles changes substantially from year to year, as the number of styles grows over time. We therefore define the *Lows* as *m* randomly chosen styles with a popularity rank higher than *m*, i.e., with $Rank(n_s(t)) > m$. This definition ensures that none of our results are artefact from finite-size fluctuations, i.e., are driven by size-variations of the data sample. All results involving *Lows* have been averaged over 1,000 random collections of low-popularity styles. For each year, we only consider styles with at least 10 releases in these groups. To ensure that our results are independent of the concrete choice of the threshold *m*, we carry out robustness tests by letting *m* vary over a wide range of choices. This test guarantees that our results are not driven by *m*, but reflect a general feature of style popularities. However, to increase the clarity of the presentation, we will present the main results for a choice of *m*.

Popularity change of a style, $\delta n_s(t)$, is measured as the percent change in popularity,

$$\delta n_s(t) = \frac{n_s(t+\delta t) - n_s(t)}{n_s(t)}. \qquad (2)$$

To identify successful styles among styles that do not yet belong to the *Highs*, styles are ranked according to their $\delta n_s(t)$ values. In the following we refer to the successful lows as *Newcomer*, i.e., as those that have a rank of $\delta n_s(t)$ of no more than *m* but which do not yet belong to the *Highs*,

$$New(t,m) = \{s \notin Highs(t,m) | Rank(\delta n_s(t)) \leq m\}. \qquad (3)$$

If not specified otherwise, we fix *m*=10 and $\delta t = 5$ years.

**Characterizing styles by instrumentation.** The relations between styles and their instruments can be encoded in a time-dependent, bipartite adjacency matrix *A*, which consists of two types of nodes, styles and instruments. If there is at least one album released at time *t* in style *s* and recorded using the instrument *i*, we set $A_{si}(t) = 1$, otherwise, $A_{si}(t) = 0$. A detailed analysis of this bipartite network can be found in [29]. Some instruments, such as vocals or guitars, are highly ubiquitous and appear in many styles, whereas other instruments are highly specific for a given style, e.g., turntables in Hip Hop. To suppress the influence of highly ubiquitous instruments, we rescale the contributions of each instrument by its inverse frequency across all styles. This procedure is highly reminiscent of the common use of inverse document frequencies as a weighting factor to increase specificity in information retrieval tasks. This gives us the weighted, time-dependent network, $M(t)$, with entries $M_{si}(t) = \frac{A_{si}(t)}{\sum_i A_{si}(t)}$. The instrumentational similarity (style similarity) between two styles $i$ at $t_1$ and $j$ at $t_2$ is defined as the cosine similarity of their instrumentation vectors,

$$\varphi_{ij}(t_1, t_2) = \frac{\sum_s M_{si}(t_1) M_{sj}(t_2)}{|M_{si}(t_1)| |M_{tj}(t_2)|}, \tag{4}$$

where $|\vec{x}|$ denotes the Euclidean norm of vector $\vec{x}$. Note that there are alternative ways to define the style similarity $\varphi_{ij}$. Particularly, we will discuss the robustness of our results with respect to the choice of a similarity measure by replacing the cosine similarity in Equation (4) by the Jaccard coefficient, by the inverse Euclidean distance between the instrumentation vectors, and by their similarity as defined by the ProbS algorithm [31, 32]. The style—style similarity network for a given time, $\varphi_{ij}(t) = \varphi_{ij}(t_1 = t, t_2 = t)$, is typically fully connected with most of the links having a relatively small weight. Such networks can efficiently be visualized by their maximum spanning tree (MST), the so-called "backbone". For a connected network (all nodes are in the giant component) with $N$ links, the MST is the set of $N$-1 links that have the largest possible weights and that span each node of the original network.

For two *sets* of styles, $S_1(t_1, m)$ and $S_2(t_2, m)$, the similarity $\Phi(S_1(t_1, m), S_2(t_2, m))$ can be defined as the average similarity of each pair of styles, where one style is chosen from $S_1(t_1, m)$ and the other from $S_2(t_2, m)$. The set-similarity $\Phi(S_1(t_1, m), S_2(t_2, m))$ is defined as,

$$\Phi(S_1(t_1, m), S_2(t_2, m)) = \frac{1}{m^2} \sum_{i \epsilon S_1(t_1,m), j \epsilon S_1(t_2,m)} \varphi_{ij}(t_1, t_2). \tag{5}$$

**Models for cultural change.** We first consider a neutral model for the evolution of musical styles. This approach is similar in spirit of the unified neutral theory of biodiversity in ecology [28]. There one seeks to explain the abundances of species (in our case: popularities of styles) using the neutrality assumption that the probability of a given species to produce offspring (i.e., release of a new album in a particular style) is proportional only to its abundance. For the evolution of musical styles, neutrality means that the popularity of a style does not depend on its structural characteristics, such as instrumentation. Such models can be formulated as Pólya-urn-like models [33]. Imagine an urn with balls of different colors. Each ball is a musical release. The color of the ball represents the musical style. At each time step there are two possible actions—a copying step and an innovation step. With probability $1 - p$, $0 \leq p \leq 1$, one draws a ball from the urn, notes its color, and puts the ball back into the urn, together with a new ball of the same color ("copying step"). With probability $p$, however, one introduces a ball with a new color ("innovation step"). The style popularity $n_s(t)$ is the number of balls with color $s$ added to the urn in a time interval $t$. Such Pólya urns are paradigmatic models for reinforcement processes [34]. On the long run there is either a co-existence of different colors in the urn (different styles) or a winner-takes-all dynamics leads to a situation where almost all balls have the same color—but no cycles. In the following, we consider modifications of this neutral model that represent the mechanisms of ST, RPT, and CDS, respectively. Each style $s$ is associated with an angle $\theta_s \in [0, 2\pi]$ that describes the orientation of the style's instrument vector, $A_{si}$. Taking the average over the angles of all balls in the urn yields the average angle, $\bar{\theta}$. There is a unique elite style $e$ characterized by $\theta_e(t)$ with popularity $n_e(t)$. The modified urn models also have a copying and innovation step, but we introduce two changes with respect to the neutral model. First, in the copying step one now either copies the current elite style or the style of a randomly chosen release. The second modification is in the innovation step. Each innovation introduces a new elite style that may or may not become successful. The parameter $p$ can be a function of the state of the urn. For ST, $p$ is proportional to the elite style popularity, $p \sim n_e(t)$ (the more people imitate an elite style, the higher are the chances for the elite to abandon that style). For RPT, $p$ is constant (new elites form at random). For CDS, $p$ is proportional to the similarity between the current elite style and the average styles of all releases, $p \sim \cos^2(\bar{\theta} - \theta_e)$ (the more uniform the current music releases are in style, the stronger the effect of CDS). In the case of RPT, the new elite style e' has a random style angle $\theta_{\acute{e}} \in [0, 2\pi]$. For ST and CDS, the new elite style is orthogonal to the current average style, $\theta_{\acute{e}} = \bar{\theta} - \frac{\pi}{2}$; see SM Text S1.

## Results

**Fashion cycles in musical styles.** Figure 2A shows fashion cycles of selected musical styles as they appear in the empirical data. For each year we identify the style with the highest popularity and show its $n_s(t)$ over the entire observation window. There is a clear pattern in which styles increase in popularity until they reach a peak, after which they start to decrease while a new style takes over. A more complete picture of the evolution of styles is shown in Figure 2B. For each year we rank styles according to their popularity, $Rank(n_s(t))$, measured as the number of releases in that style, relative to the number of all releases. Figure 2B shows a selection of those styles that ranked among the five most popular styles in at least one year during the observations. Styles typically enter at high ranks (low popularity) where they remain over a certain period. At some point, they start to increase in popularity (decrease in rank). In some cases, this occurs within two or three years. Some styles manage to maintain high popularity (low ranks) over an extended period, whereas others fade back into oblivion rapidly (e.g. Euro House after the 1990s).

**Fashion cycles in models of cultural change.** The neutral model is incapable of explaining the existence of fashion cycles as seen in Figure 2A. The modified models ST, RPT and CDS do produce such cycles; see SM Figures S1 and S2. The RPT model assumes that two consecutive high-popularity styles are independent from each other in terms of their instrumentation, while in the ST and CDS models they tend to be opposites of each other. All the proposed theories, ST, RPT, and CDS, can explain the observed fashion cycles. Further empirical tests are therefore necessary to determine which of the three competing mechanisms is at work.

**Empirical tests for theories of fashion cycles.** We now consider three different hypotheses to empirically test the theories of fashion cycles. We control the family-wise error rate (FWER) at a level of α<0.05 using the Bonferroni-Holm method. We first test if styles with high popularity indeed tend to be imitated by other styles. For this we assume that *Lows* (styles with low popularity and high popularity ranks) have an overall tendency to imitate (or follow) the most popular styles by adopting the instruments used by the *Highs* (indirectly representing the elites). Imitation can then be understood as a process that results in the similarity between *Highs* at time *t* and *Lows* at time *t* being lower than the similarity between *Highs* at *t* and *Lows* at $t + \delta t$ (*Lows* follow *Highs*). This means that *Lows* tend to "move" towards a higher similarity with the current *Highs*. Using the similarity measure $\Phi$ to describe the overlap between two sets of musical styles, we can formulate the null hypothesis to reject the process of imitation in the data,

$$H_{Imitation}: \Phi\big(Lows(t + \delta t), Highs(t)\big) \leq \Phi\big(Lows(t), Highs(t)\big). \qquad (1)$$

Figure 3A shows that the null hypothesis for imitation, $H_{Imitation}$, must be rejected given the data ($p < 0.001$, one-sided paired t-test using all observations). This means that there is a statistically highly significant effect in the data that low-popularity styles tend to imitate high-popularity styles in terms of instrumentation.

ST assumes that as a result of this imitation, elite members will adopt a new style to differentiate itself from its imitators. This new style will gain popularity and will be imitated by those that have followed the elite before. Now the style formerly adopted by the elite members will belong to the group of *Highs*, while the elite's newly adopted style belongs to the group of *Newcomers*. It follows that the similarity between newcomers at time *t* and *Highs* at time $t + \delta t$, $\Phi\big(New(t), Highs(t + \delta t)\big)$, should be larger than their similarity with the *Highs* at time *t*, $\Phi\big(New(t), Highs(t)\big)$. Otherwise this would mean that the formation of *Newcomers* is not related to concurrent changes in instrumentations of the *Highs*—in this sense *Newcomers* and *Highs* would be independent from each other. However, ST posits that the *Highs* would seek to differentiate themselves from the *Newcomers* in order to signal their distinguished status. We test the null hypothesis to reject ST,

$$H_{ST}: \Phi(New(t), Highs(t+\delta t)) \leq \Phi(New(t), Highs(t)). \qquad (2)$$

Figure 3B shows that the null hypothesis for ST, $H_{ST}$, cannot be rejected in the data (p=0.68). This result suggests that styles do not become popular because of the adoption by an elite group that is then followed by a large number of imitators. The dynamics of *Newcomers* and of the current *Highs* seems to be independent from each other, meaning that they represent different elite groups in competition with each other.

In contrast to ST, the mechanisms of RPT and CDS propose that styles become popular because of a new elite group that adopts them. In particular, if RPT were correct, one would expect that it is a randomly chosen style from the set of *Lows* that will dominate the next fashion cycle. CDS in turn would imply that the next cycle is characterized by a counter-elite that adopts an instrumentation that is in stark contrast to the current *Highs*. In other words, CDS suggests that the emerging *Newcomers* follow styles that are more dissimilar from the current *Highs* than randomly chosen *Lows*. A null hypothesis that fashion cycles can be explained on the basis of RPT instead of CDS can then be formulated as follows,

$$H_{CDS-RPT}: \Phi(New(t), Highs(t)) \geq \Phi(Lows(t), Highs(t)). \qquad (3)$$

Figure 3C shows that the null hypothesis for RPT, $H_{CDS-RPT}$, can be firmly rejected in the data ($p < 10^{-8}$). This result shows with very high significance that *Newcomers* typically represent an outsider group that tends to be in "opposition" to the current *Highs* in the use of instrumentation.

**Genre-specific empirical tests for theories of cultural change.** The three hypotheses, $H_{Imitation}$, $H_{ST}$, and $H_{CDS-RPT}$ were tested for several musical genres. We considered the three genres with the highest numbers of related styles, namely Electronic, Rock, and Folk music; see Figure 4. For Electronic, Figures 4A-C, and Rock music, Figures 4D-F, again we find that imitation is a significant feature in the data, that there is no evidence for ST, and that RPT must be rejected in the statistical test against expectations from CDS. For Folk music we find evidence for a particularly strong imitation effect, Figure 4G, and a preference of CDS over RPT, Figure 4I. However, in contrast to all other tests, we reject the null hypothesis for ST, $H_{ST}$ ($p$=0.01), see Figure 4H. This means that in Folk music we cannot rule out ST as a relevant mechanism, though statistically it is a much weaker feature of the data than when compared to CDS.

**Robustness.** There is a dependence on the parameters $\delta t$ and $m$. To ensure that our results are independent of the concrete choice of the threshold $m$, we carry out a robustness test by letting $m$ vary over a wide range of choices. This test guarantees that our results are not driven by $m$, but reflect a general feature of style popularities. There is significant evidence for an imitation effect and a preference of CDS over RPT in each test. The evidence for CDS is particularly strong over short time intervals, $\delta t$. There is no evidence for ST in almost all parameter settings, except for very small style numbers, $m$, and for long time intervals, $\delta t$; see SM Figure S3.

We also study the robustness of our main results with respect to the choice of the similarity measure for styles in Equation (4). Referring to the p-values of the three hypothesis tests shown in Figure 3, for all considered measures we find a significant imitation effect (p=0.014 for the Jaccard coefficient, p=0.024 when using the inverse

Euclidean distance and p=0.019 for ProbS), no evidence for ST (Jaccard: p=0.95, Euclid: p=0.90, ProbS: p=0.35) and a preference of CDS over RPT (Jaccard: p<10$^{-3}$, Euclid: p=0.0017, ProbS: p= p<10$^{-3}$). By comparing the results across the different similarity measures, we see that the imitation effect is a less significant feature of the data than the CDS-RPT effect. For instance, if we would control the FWER by the more conservative Bonferroni procedure (instead of the uniformly more powerful Bonferroni-Holm method), the imitation effect would disappear when using the inverse Euclidean distance or ProbS as similarity measure.

**Spreading of popularity in the network of musical styles.** Our formalism enables us to visualize the network evolution of musical styles. They relate to each other in a complex network of style-style similarities. In Figure 5 we show the backbone of the style similarity network for three time intervals of 20 years, from 1956 to 2015. Each node represents a style with a size proportional to its popularity, $n_s(t)$; the colors indicate genres. Similar styles are connected by links. Styles belonging to the same genre tend to be in close proximity to each other. For each time interval the network consists of a core of styles with high degrees, i.e., a large number of similar styles. The periphery contains low-degree nodes, i.e., styles with a low number of similar styles. Initially, styles belonging to the genres Electronic and Hip Hop correspond almost exclusively to low-popularity nodes in the periphery of the network. Over time they show large gains in popularity at the expense of former high-popularity styles lying in the core of the network, such as styles belonging to Latin music. Outsider styles in the periphery of the network can increase their chances of dominating the next fashion cycles by sending strong counter-dominance-signals. Overall, this leads to a network evolution where popularity diffuses from periphery to the core, and then dissipates to the periphery again.

## Discussion

By introducing a notion of cultural similarity, we formulated quantitative models for three possible mechanisms of cultural change, together with empirical, statistical tests to clarify which of these mechanisms describe the actual data. We use a comprehensive dataset that contains almost all major album releases since the second half of the twentieth century, including the entire range of popularity of a genre or style (i.e., from least to most popular, as opposed to other studies that focused only on popular releases, such as the Billboard Hot 100 [36]). We found that low-popularity musical styles follow and imitate stylistic characteristics of high-popularity styles in a way that cannot be accounted for by neutral models of evolution. We found very little evidence for costly signaling theory (ST), i.e., of members of an elite "fleeing" from their imitators. Instead, fashion cycles seem to emerge due to competition between members of different elite groups. Furthermore, in contrast to predictions of random pattern theories (RPT), we found that styles with characteristics in strong opposition to styles of the current elite dominate the next fashion cycles. This supports our proposed mechanism of counter-dominance-signaling (CDS) according to which changes in art and fashion styles happen whenever a new elite successfully challenges the hegemony of a previous elite. We could confirm these findings in the genre-specific tests for Electronic, Rock and Folk music, with the exception of some evidence for ST in Folk music. This indicates that ST is only a feasible mechanism of cultural change in Folk music (i.e., more traditional musical styles) over long time intervals. In conclusion, CDS drives short-term cultural change in all considered cases. A possible explanation for this finding is that folk music constitutes a genre where music is frequently used as a means of cultural identification. Research in cross-cultural psychology applicable to folk music has shown that members of cultural groups tend to choose those stylistic members that are in stark contrast to the ones endorsed by other cultural groups [14,37]. Such costly signals could help to preserve a culture's identity as they are less likely to be mimicked by members of other cultural groups.

Note that a limitation arises from our use of a partly folksonomic classification of musical styles [38]. That is, boundaries between styles emerge out of a collective and moderated effort of social tagging that might introduce some biases towards styles with a very dedicated base of users. In addition, some care needs to be taken in interpreting the results of the hypothesis tests. In particular, our rejection of RPT over CDS in the third test does not necessarily imply that CDS is the only possible explanation for the observed fashion cycles, but rather that, of all mechanisms considered here, CDS is least likely to be ruled out.

In 1905 German sociologist Georg Simmel [6] published his *Philosophie der Mode [Philosophy of Fashion]*. Still considered as one of the most influential sociological theories of the early 20[th] century, Simmel's work ignited a century-long debate about the question of whether there are specific patterns in the social dynamics that underlie art and fashion cycles and—if yes—what are the mechanisms driving them. With our research we provide an entirely data-driven and quantitative answer to this question: Cyclical changes of complex mass-cultural symbol systems manifested in art and fashion styles are driven by outside groups that successfully challenge the current elites.

Acknowledgments. PK acknowledges helpful discussions with Rudolf Hanel.

## References

1. Henshilwood C. & d'Errico F. *Homo Symbolicus: The Dawn of Language, Imagination and Spirituality.* John Benjamins Publishing Company (2011).
2. Kreuzbauer, R., King, D., & Basu, S. The Mind in the Object—Psychological valuation of materialized human expression. *Journal of Experimental Psychology: General*, 144, 764 (2015).
3. Bliege Bird, R., Smith, E. Signaling Theory, Strategic Interaction, and Symbolic Capital. *Current anthropology*, *46*(2), 221-248 (2005).
4. Rahman Khan, S. The Sociology of Elites. *Annual Review of Sociology*, *38*, 361-377 (2012).
5. Cheng, J. T., Tracy, J. L., Foulsham, T., Kingstone, A., & Henrich, J. Two Ways to the Top: Evidence that Dominance and Prestige are Distinct yet Viable Avenues to Social Rank and Influence. *Journal of Personality and Social Psychology*, *104*(1), 103 (2013).
6. Simmel, G. Fashion. *International Quarterly*, 10(1), 130-145 (1904), reprinted in *American Journal of Sociology*, *62*(6), 541-558 (1957).
7. Bourdieu, P. *Distinction: A Social Critique of the Judgement of Taste*. Cambridge, MA: Harvard University Press. (1984).
8. Aspers, P., & Godart, F. Sociology of Fashion: Order and Change. *Annual Review of Sociology*, 39, 171-192 (2013).
9. Acerbi A, Ghirlanda S, Enquist M. The Logic of Fashion Cycles. *PLoS ONE* 7(3): e32541. https://doi.org/10.1371/journal.pone.0032541 (2012).
10. Chwe, M. S. Y. *Rational Ritual: Culture, Coordination, and Common Knowledge*. Princeton University Press (2013).
11. Turchin, P. *Ages of Discord*. Beresta Books (2016).
12. Pesendorfer, W. Design Innovation and Fashion Cycles. *The American Economic Review*, 771-79 (1995).
13. Veblen, T. *The Theory of the Leisure Class*. Oxford University Press (1899/2009).
14. Berger, J. & Ward M. Subtle Signals of In-conspicuous Consumption. *Journal of Consumer Research*, 37 (4), 555–69 (2010).
15. Blumer, H. Fashion: From Class Differentiation to Collective Selection. *The Sociological Quarterly*, *10*(3), 275-291 (1969).
16. Kaiser S., Nagasawa R. & Hutton S. Construction of an SI theory of fashion: Part 1. ambivalence and change. *Clothing and Textiles Research Journal.* Vol 13, Issue 3, pp. 172 – 183 (1995).
17. Crane D. Diffusion Models and Fashion: A Reassessment. *The ANNALS of the American Academy of Political and Social Science.* Vol 566, Issue 1, pp. 13 – 24 (1999).
18. Tschmuck, P. *Creativity and Innovation in the Music Industry*. Springer (2012).
19. Stanley B. *Yeah Yeah Yeah: The Story of Modern Pop*. London: Faber & Faber (2013).
20. Peterson RA, Berger DG. Cycles in Symbol Production. *Am. Soc. Rev.* 40, 158–173. (doi:10. 2307/2094343) (1975).

21. Bentley, R.A., Hahn, M.W. and Shennan, S.J., 2004. Random Drift and Culture Change. *Proceedings of the Royal Society of London B: Biological Sciences*, 271(1547), pp.1443-1450.
22. Bentley, R.A., Lipo, C.P., Herzog, H.A. and Hahn, M.W., 2007. Regular Rates of Popular Culture Change Reflect Random Copying. *Evolution and Human Behavior*, 28(3), pp.151-158.
23. Yoganarasimhan, H., 2017. Identifying the Presence and Cause of Fashion Cycles in Data. *Journal of Marketing Research*, 54(1), pp.5-26.
24. Kreuzbauer R. & Cheong B. Strategies of Counterdominance – When Luxury Doesn't Give You Power, *Proceedings of the Society for Consumer Psychology Conference*, Vienna (2015).
25. Boehm, C. Egalitarian Behavior and Reverse Dominance Hierarchy. *Current Anthropology*, *34*(3), 227-254 (1993).
26. Schelling, T. C. *The Strategy of Conflict*. Harvard University Press (1980).
27. Muthukrishna, M., & Henrich, J. Innovation in the Collective Brain. *Phil. Trans. R. Soc. B*, *371*(1690), 20150192 (2016).
28. Hubbell, SP. *The Unified Neutral Theory of Biodiversity and Biogeography*. Princeton University Press (2001).
29. Percino G, Klimek P, Thurner S. Instrumentational Complexity of Music Genres and Why Simplicity Sells. *PLoS ONE* 9(12): e115255. https://doi.org/10.1371/journal.pone.0115255 (2014).
30. http://www.discogs.com, retrieved February 2016.
31. Zhou T, Kuscsik Z, Liu J-G, Medo, M, Wakeling JR, Zhang Y-C. Solving the apparent diversity-accuracy dilemma of recommender systems. Proceedings of the National Academy of Sciences of the United States of America 107(10): 4511-4515 (2010).
32. Yildirim MA, Coscia M. Using random walks to generate associations between objects. PLoS ONE 9(8): e104813. https://doi.org/10.1371/journal.pone.0104813
33. Hoppe FM. Pólya-like Urns and the Ewens' Sampling Formula. *Journal of Mathematical Biology* 20(1): 91-94 (1984).
34. Eggenberger F, Pólya G. Über die Statistik verketteter Vorgänge. *Journal of Applied Mathematics and Mechanics* 3(4): 279-89 (1923).
35. Holm S, A simple sequentially rejective multiple test procedure. *Scandinavian Journal of Statistics* 65-70 (1979).
36. Mauch M, MacCallum RM, Levy M, Leroi AM. The evolution of popular music: USA 1960-2010. *Royal Society Open Science*, 2, 150081 (2015).
37. Kreuzbauer R, Chiu CY, Lin S, Bae SH. When Does Life Satisfaction Accompany Relational Identity Signaling: A Cross-Cultural Analysis. *Journal of Cross-Cultural Psychology*, 45 (4), 646-659, (2014)
38. Gruber T. Ontology of Folksonomy: A Mash-up of Apples and Oranges. *Journal on Semantic Web and Information Systems (IJSWIS)* 3(1): 1-11 (2007).

Figures

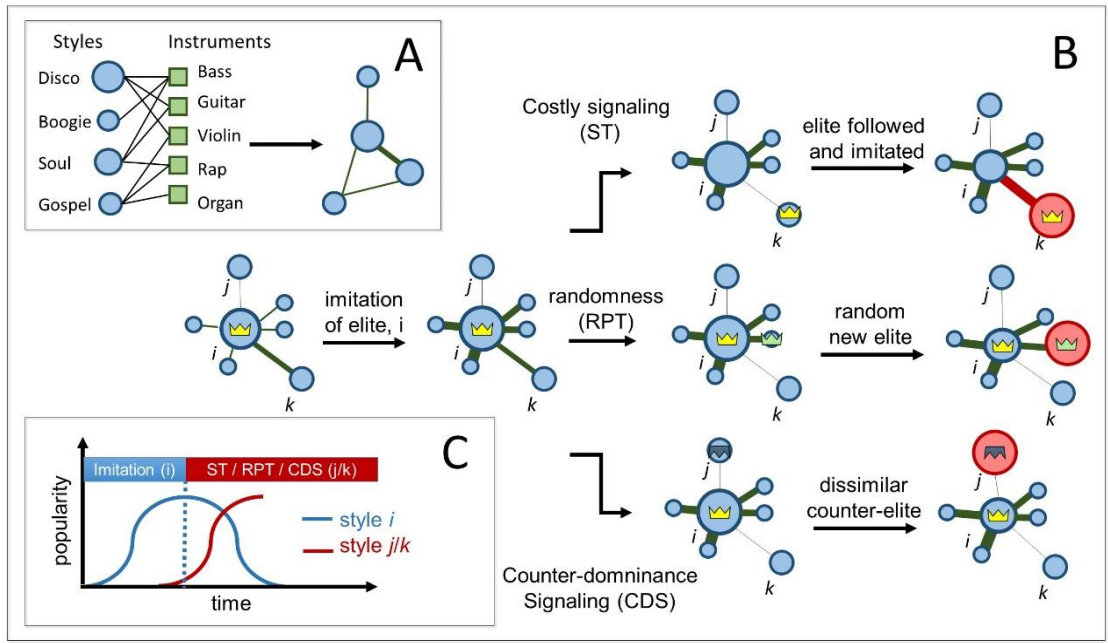

**Figure 1:** Network evolution of competing theories of cultural change. (A) The characteristics of each musical style (blue circles) are given by the instruments that are typically associated with this style (green squares). The similarities of two musical styles are measured by the number of instruments they share, leading to a style-style similarity network. The size of the circles is proportional to their popularity, the thickness of the link connecting two styles is proportional to their similarity. (B) Competing theories of cultural change imply different types of evolution of the network of musical styles. We consider a network with an elite (yellow crown) that initially adheres to style i. The popular style i will be imitated by other styles (links to i increase in thickness). Following costly signaling theory (ST), the elite seeks to differentiate itself from imitators and adopts a new style, k. Random pattern theory (RPT) suggests that a new elite (green crown) will emerge at a random position in the network. Counter-dominance signaling (CDS) predicts the emergence of a new counter-elite (blue, upside-down crown) that is highly dissimilar from the current elite, shown here for style j. (C) All three theories, ST, RPT, and CDS, give rise to fashion cycles in which style i initially increases in popularity under imitation by other styles until a new style emerges through ST, RPR, or CDS, and then dominates the next fashion cycle.

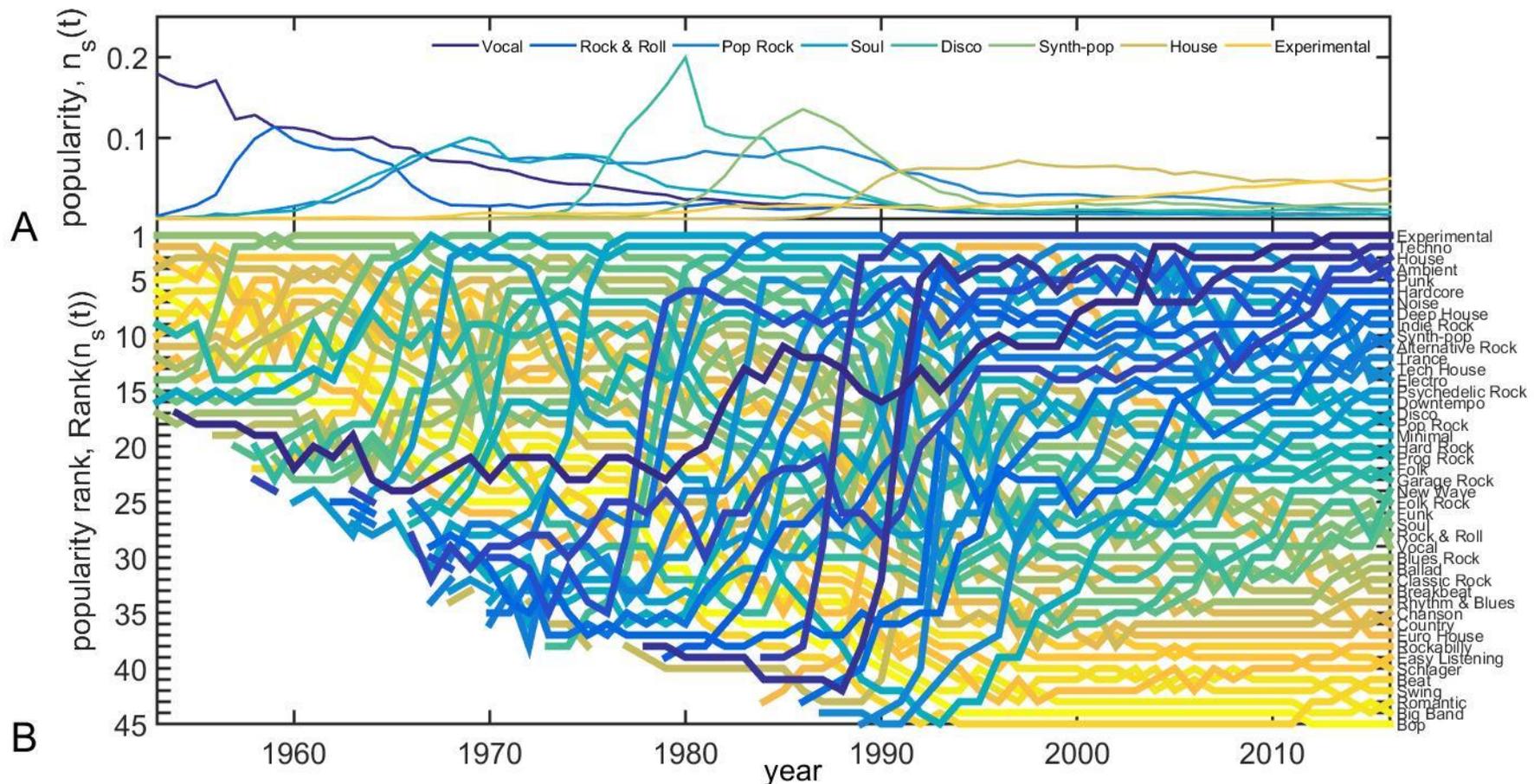

**Figure 2:** Fashion cycles in musical styles as ranked by their popularity $n_s(t)$. (A) For each year we identify the most popular style and show its popularity $n_s(t)$ over the entire observation period. The first cycle is given by Vocal (music strong focus on voice), followed by Rock&Roll, Pop Rock, Soul, Disco, Synth-pop, House, and finally experimental music. (B) Most styles enter with a rather low popularity (high rank) that they maintain over a number of years. These phases are eventually followed by a rapid increase in popularity (decrease in rank). Here we show only styles that were among the five most popular ones in at least one year.

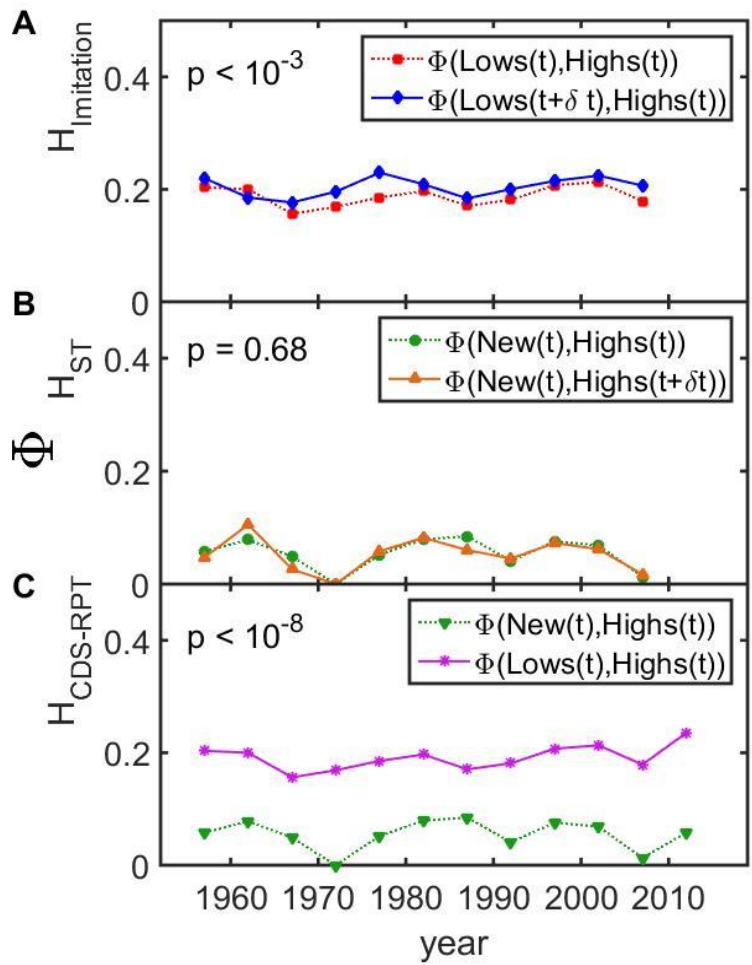

**Figure 3:** Three hypothesis tests for theories of cultural change. In each panel a test represents where the null hypothesis is rejected if the similarities shown by solid lines are significantly larger than the similarities shown by dotted lines. (A) An imitation effect in the data implies that Lows have a tendency to become more similar to the Highs over time. The corresponding null hypothesis can be rejected, i.e., there is significant imitation in the data. (B) Under ST, one would expect that Newcomers emerge from stylistic changes of the current Highs. The corresponding null hypothesis cannot be rejected, i.e., there is no evidence for ST. (C) As opposed to RPT, CDS suggests that the next fashion cycle will be dominated by a counter-elite through its use of stylistic elements that are in direct opposition to the current elite. The corresponding null hypothesis can be rejected in the data. Cultural change occurs through counter-dominance signals and not by random choices.

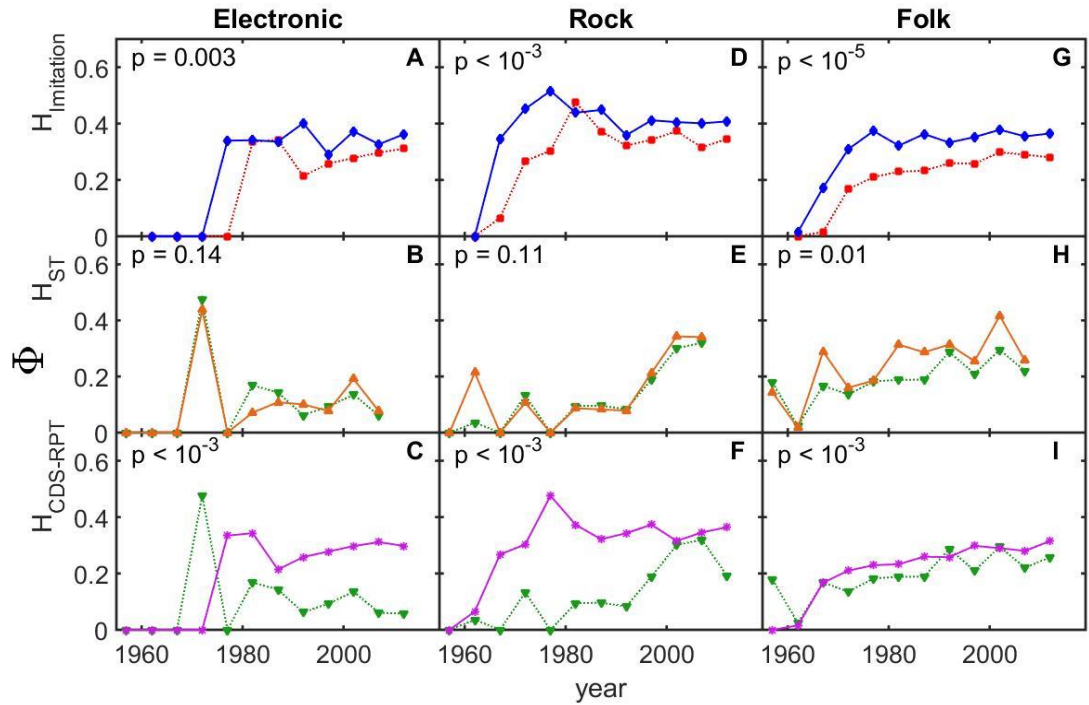

**Figure 4:** Genre-specific tests for three theories of cultural change. For Electronic and Rock music we observe a significant imitation effect, no evidence for ST, and a preference of CDS over RPT. For Folk music we observe a particularly strong imitation effect, and evidence for ST and CDS.

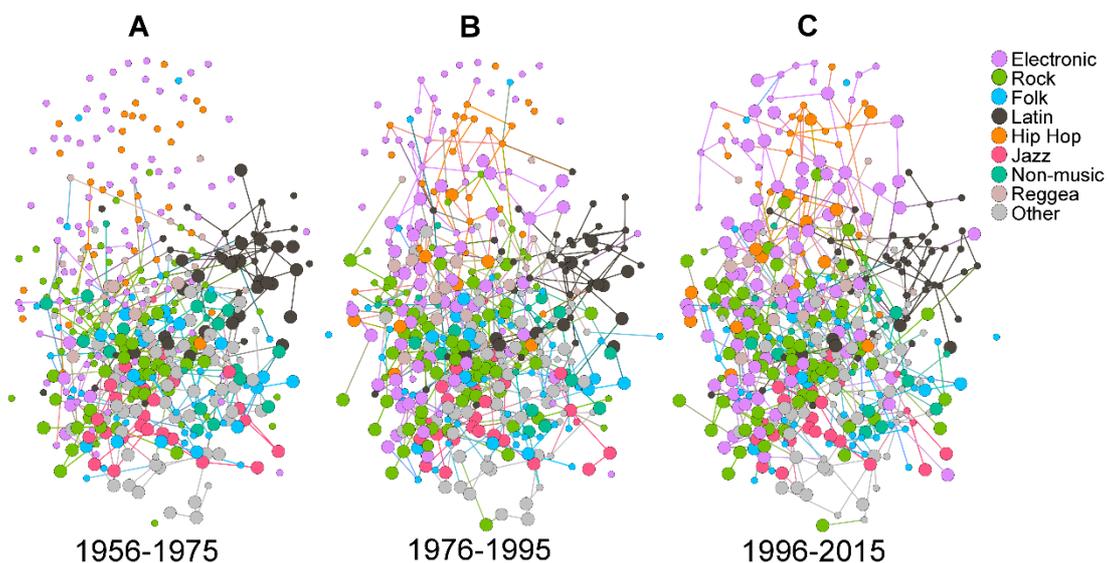

**Figure 5:** Dynamics of the style-style similarity network. We show the MST of the style-style similarity network for three time intervals. Nodes correspond to styles with colors given by their genre. The size of the nodes is proportional to their popularity $n_s(t)$. There is a large number of styles in the periphery of the network with substantial gains in popularity, whereas some styles in the core of the network decrease in popularity, e.g., Latin music styles.

# Supplementary Material

## Text S1

**Models for cultural change.** In the model for ST, CDS, and RPT, the copying step is modified as follows. The model parameter $\mu$, $0 \leq \mu \leq 1$, is introduced that gives the probability of an artist adopting the current elite style when producing a musical release. With probability $1 - \mu$ the copying step is carried out exactly as specified before.

The neutral model for cultural change is initialized with the number of balls (musical releases) and their colors (styles) being taken from the data for 1956. The initial state of the urn at $t = 0$ is given by 2,283 balls with 123 different colors. The probability to draw a ball of color $s$ is given by $n_s(t = 0)$. To match the time scale of the model with the data, we count the number of musical releases in each year $t$, $N(t)$. For each year $t$ we repeat the process described below until we have added $N(t)$ new balls, after which we advance to time $t+1$. The neutral model is then specified as follows, starting at time $t$=1.

1. **Copying step.** With probability 1-$p$, perform a *copying step*. Randomly draw a ball that has been added at time $t$-1. Place it back into the urn, along with a new ball of the same color.
2. **Innovation step.** Otherwise, with probability $p$, perform the *innovation step*. Add a ball to the urn with a *new* color.
3. Repeat the copying or innovation step until $N(t)$ balls have been added; then proceed to time $t$+1.

For the ST, RPT, and CDS models we assume that each style $s$ is characterized by a style angle $\theta_s$. At each point in time there is a unique elite style $e$ with angle $\theta_e$ and popularity $n_e(t)$. Initially, at $t$=0, the elite style is identified with the style with the highest popularity in the data and all styles are assigned a random angle $\theta_s \in [0,2\pi]$. Denote by $\bar{\theta}(t)$ the average style of all balls that have been added at time $t$. The ST, RPT, and CDS models introduce two modifications to the neutral model specified above.

- **Modification of the copying step.** With probability $1 - \mu$, copy the style of the ball that has been drawn. Otherwise, with probability $\mu$, copy the elite style.
- **Modification of the innovation step.** The probability, $p$, that a new elite style is introduced can now be a function of the state of the urn. We consider three ways to specify this function.
    - **ST**: with probability $p = c\, n_e(t)$ introduce a new elite style $e'$ with style angle $\theta_{\acute{e}} = \bar{\theta}(t) - \frac{\pi}{2}$, $c$ being a constant.
    - **RPT**: with probability $p = c$ introduce a new elite style $e'$ with a style angle chosen from the uniform distribution over $[0,2\pi]$.
    - **CDS**: with probability $p = c \cos^2(\bar{\theta}(t) - \theta_e)$ introduce a new elite style $e'$ with style angle $\theta_{\acute{e}} = \bar{\theta}(t) - \frac{\pi}{2}$.

Note the differences between RPT and the neutral model: whereas the neutral model assumes that changes in the popularity of styles are driven by the same, uncorrelated, and random effects for each style, RPT assumes that these effects are different from each other and determined by seemingly random external factors that are not represented in the instrumentation data.

Results for all models are shown in the Supporting Figures 1 ($\mu = 0.1$) and 2 ($\mu = 0.9$) for $c = 10^{-5}$. As was done in Figure 2A for the data, we identify for each year the most popular style and show its popularity from 1956 onwards. In addition, for the ST, RPT, and CDS models, we show the "direction" of styles using arrows in matching colors rotated by $2\theta_e \mod(2\pi)$ degrees (so that orthogonal styles are marked with arrows that point in opposite directions). We observe no fashion cycles for the neutral model in either case. The higher are the values of $\mu$, the slower are the growth and decline rates for individual styles. For RPT we see a random sequence of high-popularity styles. The duration of the cycles appears to be more erratic than for ST and CDS, meaning that styles are sometimes replaced at lower popularity levels under RPT. ST and CDS produce very similar patterns. For $\mu = 0.1$ two consecutive high-popularity styles often are complete opposites of each other (orthogonal style angles), while for $\mu = 0.9$ this effect is less pronounced. Note that it is often the case in all three models that a new elite style is introduced and does *not* become the most popular style before it is replaced by a new elite style. Nevertheless, two consecutive high-popularity styles seem to appear as opposites of each other.

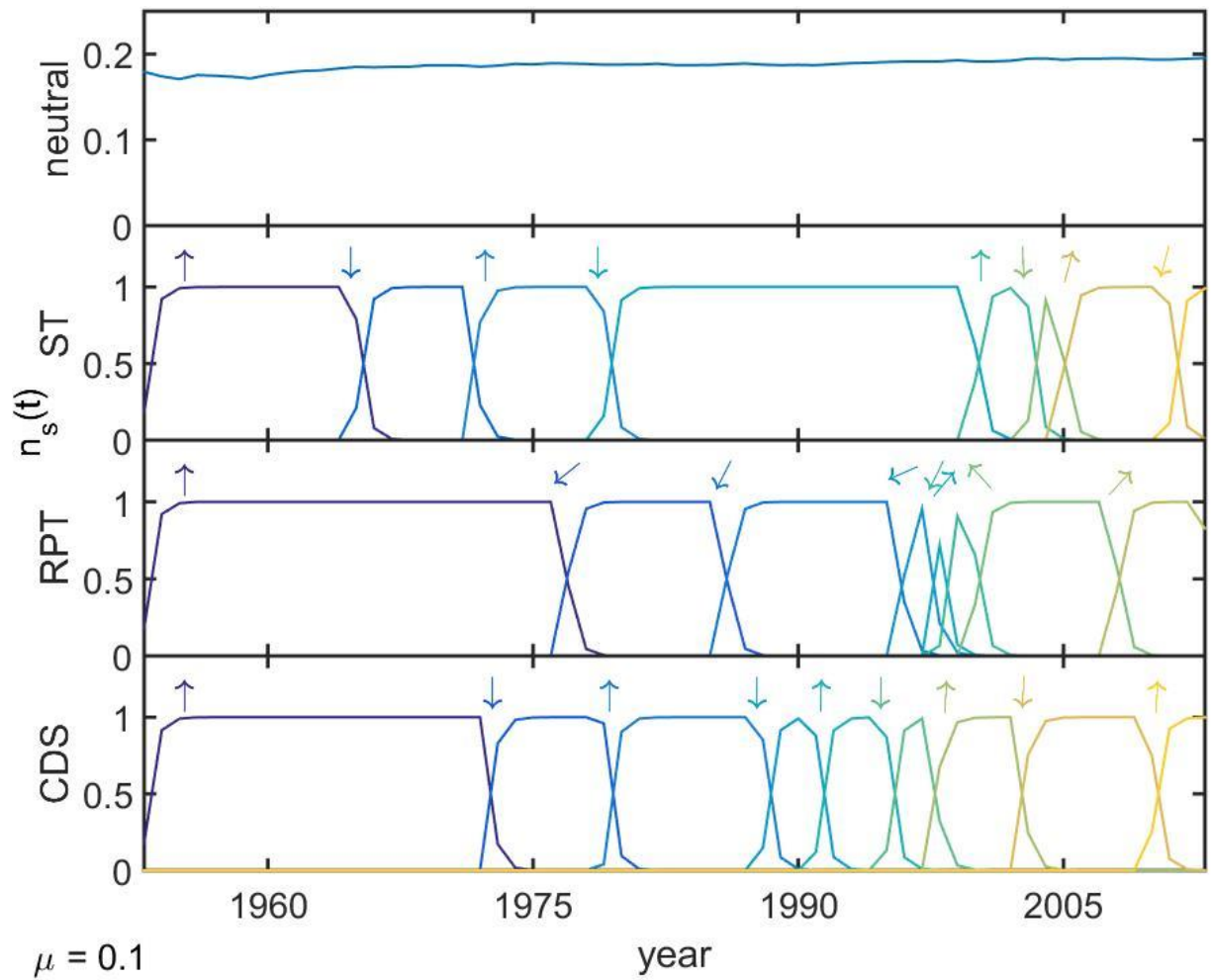

**Supporting Figure S1:** Dynamics of the most popular styles in the Pólya-urn-like models for fashion cycles for $\mu = 0.1$ and $c = 10^{-5}$. There are no cycles in the neutral model. For ST, RPT, and CDS, cycles do appear. Styles are characterized by angles which are shown using arrows in matching colors for each high-popularity style. Arrows that point in the same direction indicate that the corresponding styles have the same angle. Arrows pointing in opposite direction indicate that their styles are opposites of each other. For RPT we see a random sequence of style angles with quite erratic durations. For ST and CDS we observe that consecutive high-popularity styles appear as opposites of each other.

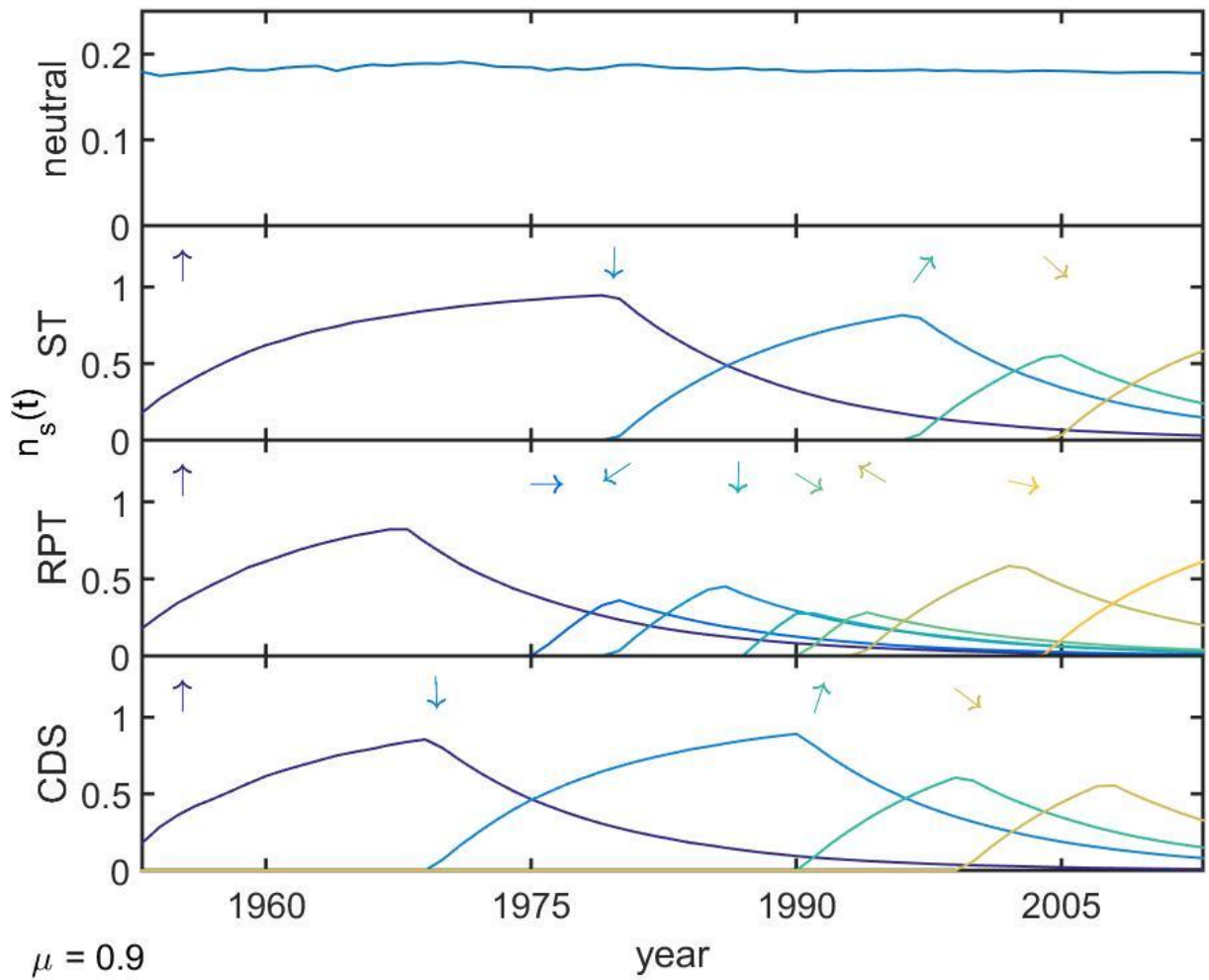

**Supporting Figure S2:** Same as Supporting Figure S1 for $\mu = 0.1$. The cycles in ST, RPT, and CDS have slower growth and decline rates as compared to higher values of $\mu$. Stylistic differences between consecutive high-popularity styles in ST and CDS are also not as pronounced as seen before.

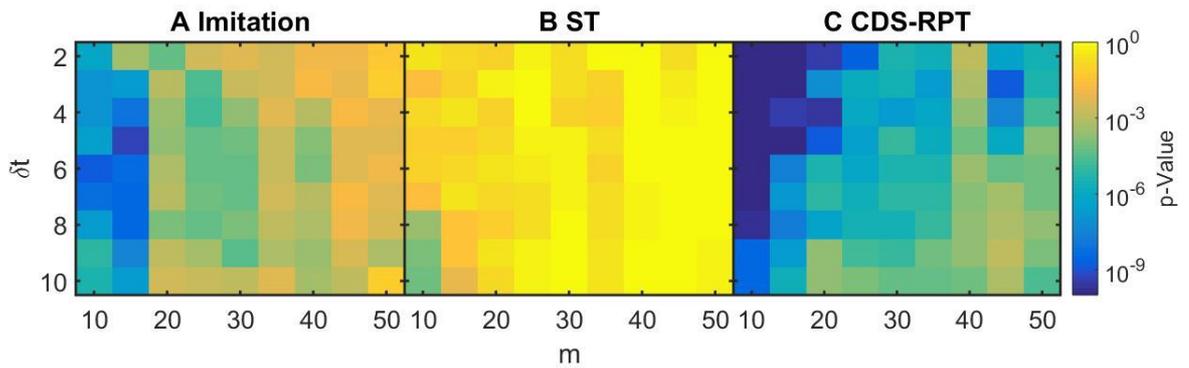

**Supporting Figure S3:** Robustness tests for the three hypothesis tests for theories of cultural change. For (A) $H_{Imitation}$, (B) $H_{CST}$, and (C) $H_{CDS-RPT}$ we repeat the hypothesis tests for different values of m and $\delta t$. The colors indicate the corresponding p-values on a scale from insignificant (yellow) to highly significant (dark blue). Imitation effects and CDS are confirmed at high levels of significance for all considered parameter settings. The ST hypothesis can be rejected for almost all of the considered settings, with some exceptions for very long time intervals $\delta t$ and very low numbers of musical styles in the test sets, m.